\input{aipcheck}

\documentclass[
    ,final            
  ]
  {aipproc}
\layoutstyle{8x11double}

\begin{document}

\title{Improved Measurement of Electron Antineutrino Disappearance at Daya Bay}


\author{Xin Qian~\footnote{Email:xqian@caltech.edu}, on behalf of the Daya Bay Collaboration}{
  address={Kellogg Radiation Laboratory, California Institute of Technology, Pasadena, CA}
}

\begin{abstract}
The Daya Bay experiment was designed to be the largest and the deepest 
underground among the many current-generation reactor antineutrino 
experiments. With functionally identical detectors deployed at multiple baselines, 
the experiment aims to achieve the most precise measurement of 
$\sin^2 2\theta_{13}$. The antineutrino rates measured in the two near 
experimental halls are used to predict the rate at the far experimental 
hall (average distance of 1648 m from the reactors), assuming there is 
no neutrino oscillation. The ratio of the measured over the predicted far-hall
 antineutrino rate is then used to constrain the $\sin^2 2\theta_{13}$.  
The relative systematic uncertainty on this ratio 
is expected to be 0.2$\sim$0.4\%.  In this talk, we present an improved 
measurement of the electron antineutrino disappearance at Daya Bay. 
With data of 139 days, the deficit of the antineutrino rate 
in the far experimental hall was measured to be 0.056 $\pm$ 0.007 (stat.) 
$\pm$ 0.003 (sys.). In the standard three-neutrino framework, the 
$\sin^2 2 \theta_{13}$ was determined to be 
0.089 $\pm$ 0.011 at the 68\% confidence level in a rate-only analysis. 
\end{abstract}

\maketitle

\section{Introduction}

As fundamental particles in the standard model, (anti)neutrinos were initially 
thought to have zero mass. Such an assumption was supported by the experimental 
evidence that only left-handed neutrinos (also right-handed antineutrinos) 
were detected~\cite{helicity}. However, in the past decades, the phenomenon 
of neutrino flavor oscillation observed by Super-K, SNO, KamLAND, MINOS, 
and many other experiments successfully established the existence of non-zero 
neutrino masses and the neutrino mixing. A recent review can be found in 
Ref.~\cite{mckeown_review}. The neutrino oscillations are commonly
described by the Pontecorvo-Maki-Nakagawa-Sakata (PMNS) matrix and two 
neutrino mass-squared differences ($\Delta m^2_{32}:=m^2_{3}-m^2_{2}$ and 
$\Delta m^2_{21}:=m^2_{2}-m^2_{1}$)~\cite{ponte1,ponte2,Maki}.
The PMNS matrix denotes the mixing between the neutrino flavor and mass 
eigenstates. It contains three mixing angles $\theta_{12}$, $\theta_{23}$, 
and $\theta_{13}$, and an imaginary phase $\delta$, referred to as the 
CP phase in the leptonic sector. 

As of two year ago, $\theta_{13}$ was still the least known among all 
three neutrino mixing angles in the PMNS matrix. The 
best constraint was from CHOOZ reactor antineutrino experiment with 
$\sin^2 2\theta_{13} < 0.17$ at 90\% confidence level (C.L.)~\cite{chooz1,chooz2}. 
A global analysis~\cite{Fogli} in 2008 including both solar and 
reactor neutrino data suggested a non-zero $\theta_{13}$. 
However, in the past 18 months or so, an explosion of data from multiple
experiments greatly enhanced our understanding of $\theta_{13}$. 
In 2011, through measurements of $\nu_e$ appearance from 
a $\nu_{\mu}$ beam, the long baseline experiments T2K~\cite{T2k} 
and MINOS~\cite{MINOS} reported hints of a non-zero $\theta_{13}$ at 
about 2.5 and 1.4 standard deviations~\footnote{Results from the long 
baseline experiments actually depend on the assumption of the neutrino 
mass hierarchy and value of CP phase $\delta$.}, respectively. 
In January 2012, the reactor antineutrino experiment 
Double-CHOOZ~\cite{Dchooz} also reported a hint of a non-zero $\theta_{13}$ 
at 1.6 standard deviations with a single detector. The Daya Bay experiment,
with six functionally identical detectors at three locations, carried out a 
measurement of relative ratio of reactor antineutrino rates~\cite{ratio}, 
which significantly improved the sensitivity to the $\sin^22\theta_{13}$. 
In March 2012, the Daya Bay collaboration announced a non-zero value of $\theta_{13}$ 
at 5.2 standard deviations~\cite{dyb_prl}. About one month later, this 
finding was confirmed by the RENO reactor antineutrino experiment~\cite{reno}, 
which reported a consistent result using a ratio between antineutrino rates 
from two detectors. The Daya Bay experiment has since reported an improved 
measurement of the electron antineutrino disappearance~\cite{dyb_cpc} 
with 2.5 times of the previously reported statistics~\cite{dyb_prl}. 

\section{The Daya Bay Experiment}

\begin{figure}[]
\centering
\includegraphics[width=75mm]{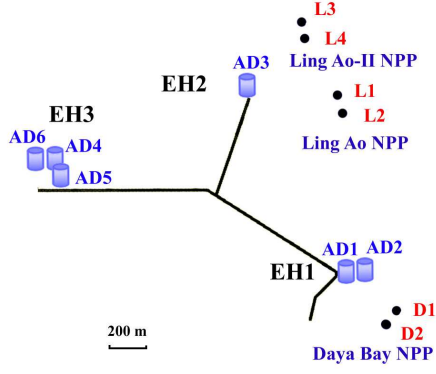}
\caption{The layout of the Daya Bay experiment. The dots represent reactor 
cores, labeled as D1, D2, and L1-4. Six ADs were installed in three experimental halls during 
the reported analysis period.}
\label{fig:layout}
\end{figure}

The Daya Bay experiment,  located on the south coast of China 
(55 km northeast to Hong Kong and 45 km east to Shenzhen), was designed to 
provide the most precise measurement of $\theta_{13}$ 
with a sensitivity of $\sin^22\theta_{13}<0.01$ 
at a 90\% C.L.~\cite{dyb_proposal}. 
Such a measurement requires high accuracy and precision. 
The high accuracy is achieved by the combination of powerful reactors 
(17.6 GW thermal power) and large target mass (80 tons in the far hall). 
In addition, the location of the far detectors is optimized to obtain the 
best sensitivity to $\sin^22\theta_{13}$ with the current knowledge of 
$\Delta m^2_{32}$. To achieve high precision, the reactor-related 
systematic uncertainties are minimized by adapting the ratio 
method~\cite{ratio} with multiple detectors at multiple 
baselines. The detector-related systematic uncertainties are minimized by 
using identical detectors and performing precise detector calibrations. 
The background-related systematic uncertainties are minimized by placing 
detectors deep underground in order to reduce cosmic muon related backgrounds. 
Furthermore, passive shielding (water pools surrounding detectors) and active 
shielding (resistive-plate chambers above water pools) were implemented to 
tag the cosmic muons. The water pools also shield detectors from 
various environmental radioactive backgrounds. 

Fig.~\ref{fig:layout} shows the layout of the Daya Bay experiment 
during this reported analysis period. There are six reactors grouped into 
three pairs. Each reactor contains a core with a maximum 2.9 GW thermal power. 
Three underground experimental halls (EHs) are connected with horizontal 
tunnels. The effective vertical overburdens are 250, 265, and 860 water-equivalent 
meters for EH1, EH2, and EH3, respectively. For this improved measurement, 
two, one, and three antineutrino detectors (ADs) were installed in EH1, EH2, 
and EH3, respectively. The distance from the six ADs to the six cores were
surveyed with the Global Positioning System (GPS) above ground and 
Total Stations underground. The precision of distance was about 
1.8 cm.  

\begin{figure}[]
\centering
\includegraphics[width=75mm]{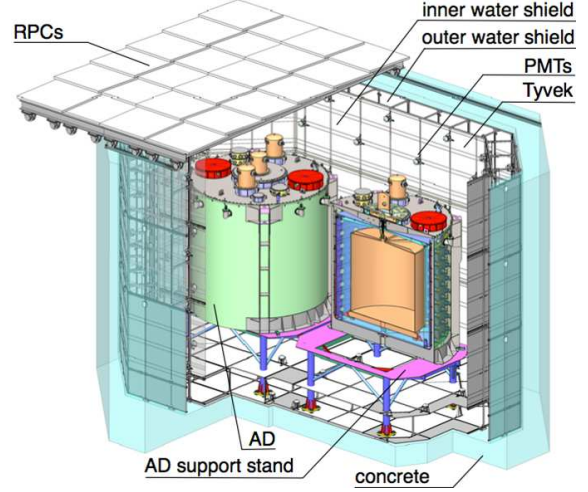}
\caption{Layout of Daya Bay detectors in a near site.}
\label{fig:detector}
\end{figure}

\begin{figure}[]
\centering
\includegraphics[width=120mm]{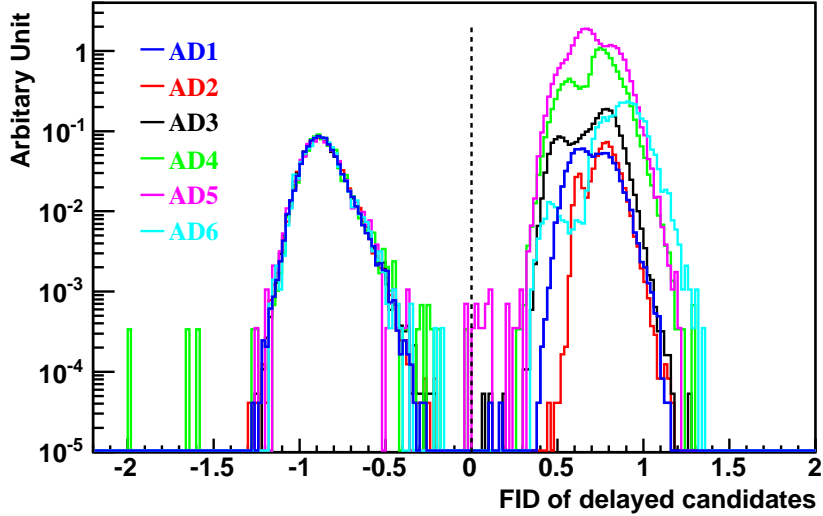}
\caption{Discrimination of the flasher events (FID $>0$) from the 
delayed signals of IBD (FID $<0$). While the flasher events possess 
different distributions among different ADs, the delayed signals of 
IBD share the same distribution. }
\label{fig:flasher}
\end{figure}

The antineutrinos in the Daya Bay experiment are detected through the 
inverse beta decay (IBD) process $\bar{\nu}_e + p \rightarrow e^+ + n$. 
The signature of such process is a prompt signal of the scintillation and 
subsequent annihilation of the position in the liquid scintillator (LS),
followed shortly by a delayed signal with $\sim$8 MeV energy deposition 
when the neutron is captured by the Gadolinium (Gd) doped 
inside LS (0.1\% in weight). 
The energy of neutrino can be deduced from the energy deposition 
of the prompt signal with $E_{\bar{\nu}} \approx E_{positron} + 0.8{\rm~MeV}$. 
As shown in Fig.~\ref{fig:detector}, the ADs adopt a three-zone 
cylindrical-shaped design, with the inner, middle, and outer zone containing 
20 ton Gd-doped LS (Gd-LS), 20 ton LS, and 40 ton mineral oil, respectively. 
With load cells and an ISO tank, the target mass uncertainty of the 
20 ton GD-LS is controlled to be only about 3 kg.

Each AD contains 192 8-inch photomultiplier tubes (PMTs) 
installed on the side walls. The photocathode coverage is about 8\%, 
which is further enhanced to about 12\% with a pair of optical reflectors 
at the top and bottom of each detector. The achieved detector energy 
resolution is parametrized as~\cite{dyb_nim} 
\begin{equation}
\frac{\delta E}{E} = (\frac{7.5}{\sqrt{E({\rm MeV})}}+0.9)\%,
\end{equation}
with respect to the visible energy $E$. The detector calibration is performed 
weekly with three automated calibration units (ACUs) per AD: two located 
above the center and the edge of the GD-LS region and one placed above 
the LS region. The ACUs are remotely controlled by a LabVIEW program and 
operated underwater. Each ACU contains four sources: a light-emitting diode (LED) 
for the PMT gain/timing calibration, a $\sim$15 Hz $^{68}$Ge source for the threshold 
calibration of the IBD prompt signal, a $\sim$100 Hz $^{60}$Co source for the high statistical 
determination of the overall energy scale, and a $\sim$0.5 Hz $^{241}$Am-$^{13}$C 
neutron source to understand neutron captures on Gd and to determine the 
H/Gd ratio in the target region.

The muon detection system in each experimental hall consists of a high purity water pool
surrounding the ADs and a layer of resistive plate chambers (RPC) above the 
water pool. The water pool is further divided into two optically separated regions, 
the inner water pool (IWS) and the outer water pool (OWS). Each region operates as 
an independent water Cerenkov detector, and can be used to cross calibrate each other. 
The muon detection efficiencies are measured to be 99.7\% and 97\% for the IWS and
 OWS~\cite{dyb_nim}, respectively.
The water pool also plays a crucial role in shielding radioactive backgrounds. 
The distance between the edge of each AD to the closest wall is at least 2.5 m.  
The array of RPCs covering the entire water pool is used to provide additional tagging of 
cosmic muons. 

\section{Selection of Inverse Beta Decay Events}

About 5\% of the PMTs would spontaneously flash and emit light. Such 
events are called ``flashers''. The reconstructed 
energy of such events covers a wide range, from sub-MeV to 100 MeV. A few 
features are observed when a PMT flashes: i) the charge fraction of 
the flashing PMT is high; 
ii) the surrounding PMTs as well as the ones located on the opposite 
side of the AD receive large fraction of light from the flashing PMT; 
and iii) the timing spread of all PMTs' hits are generally wide. 
Accordingly, 
a few flasher identification (FID) variables  were constructed to separate 
the good physics events from the flashing events. 
Fig.~\ref{fig:flasher} shows the distribution of an 
FID variable deduced from the charge pattern for the IBD delay candidates.
The good IBD events are well separated from the flasher events. 
Detailed description of the flasher discrimination can be found in 
Ref.~\cite{dyb_nim}. The inefficiency of 
the IBD signals due to FID cuts is only about (0.024 $\pm$ 0.006)\%, 
and the contamination of flashing events in the IBD sample is below 
0.01\%, which is further suppressed by the accidental background subtraction
procedure. 

After the flashing events are removed, the IBDs are further selected with the following 
cuts: i) the energy of the prompt signal is between 0.7 and 12 MeV; 
ii) the energy of the delay signal is between 6 and 12 MeV; 
and iii) the time difference between the prompt and the delay signal is between
1 and 200 $\mu$s. In addition, a multiplicity cut is applied to remove the energy 
ambiguities in the prompt signal. For example, one choice of 
the multiplicity cut requires no prompt-like signal 400 $\mu$s before the delay signal and no 
delay-like signal 200 $\mu$s after the delay signal~\footnote{The prompt-like
and delay-like signals refer to events with energy 0.7-12 MeV and 6-12 MeV, 
respectively.}. The fixed time cut (relative to the delay signal) leads to 
simplified calculations of the efficiency and livetime of IBD events and 
 the rate of accidental backgrounds. Correspondingly, three types of muon veto cuts
are also applied to the delay signal in order to suppress backgrounds.
The first one is for the water pool muon, which is defined as one 
IWS or OWS event with more than 12 PMT hits. 
The veto cut spans from 2 $\mu$s before to 600 $\mu$s 
after the water pool signal. The second one is for the AD shower muon,
which is defined as one AD signal with more than 3$\times$10$^5$ photoelectrons 
(PEs). The corresponding energy is about 1.8 GeV. The veto cut spans
from 2 $\mu$s before to 0.4 s after the AD signal.
The last one is for the AD non-shower muon, which is defined as one AD signal 
with its energy between 20 MeV and 1.8 GeV. The veto cut spans
from 2 $\mu$s before to 1.4 ms after the AD signal. 
The choice of 2 $\mu$s before the muon signal in all three cuts is to leave enough room for 
the potential non-synchronization among  different detectors. 
The AD shower muon veto is applied to suppress the long-lived $^9$Li/$^8$He
background. The AD non-shower muon veto is applied to suppress
the fast neutron background. 
The overall detection efficiency for the IBD 
events is about 80\%, with the Gd catpure ratio (84\%) and the 
efficiency of the 6 MeV delay signal cut (91\%) being the two leading
contributors.

\section{Summary of Backgrounds}

The largest contamination in the Daya Bay IBD sample is the accidental 
background with about 4.6\% at the far site and about 1.7\% at the near sites. 
An accidental background event arises when 
a delay-like signal and an unrelated prompt-like signal (usually radioactive events) 
accidentally fall within a 199 $\mu$s coincidence window. 
Such background ($R_{accidental}$) can be accurately 
calculated through Poisson statistics given the measured rates of single 
prompt-like events ($R_p$) and single delay-like events ($R_d$)~\footnote{The 
efficiency due to the fixed-time multiplicity cut is then 
$P(0,400\mu s \cdot R_p)\cdot P(0,200\mu s \cdot R_d)$.}:
\begin{eqnarray}
R_{accidental} &=& P(0,200\mu s \cdot R_p)\cdot P(1,199\mu s \cdot R_p) \nonumber \\
	& & \cdot R_d \cdot P(0,200\mu s \cdot R_d).
\end{eqnarray}
Poisson function $P(n,\mu) = e^{-\mu} \frac{\mu^n}{n!}$ represents the probability of
observing $n$ events given an expectation value of $\mu$ events. 
The above calculation results in negligible systematic uncertainties. 
This method is cross-checked with an off-window coincidence method and a 
coincidence vertex method~\cite{dyb_cpc}.

The second largest contamination at the far site 
is the correlated backgrounds from the 
Am-C neutron source. During the data taking, the Am-C
neutron sources are parked inside the ACUs on top of the ADs.
The energetic neutrons from these sources occasionally undergo 
an inelastic scattering with an iron nuclei resulting in gamma emissions, 
followed by a neutron capture on another iron nuclei with additional 
gamma emissions. When these gammas are emitted toward AD, it is 
possible these correlated events could mimic an IBD signal. 
The contamination from the Am-C source, which is about 0.3\% 
(0.03\%) at the far (near) site, is calculated through a GEANT4-based 
Monte Carlo simulation, which can reproduce the energy spectrum
of single backgrounds from Am-C sources. The systematic uncertainty
of this correlated background is assumed to be 100\%.

\begin{figure}[]
\centering
\includegraphics[width=75mm]{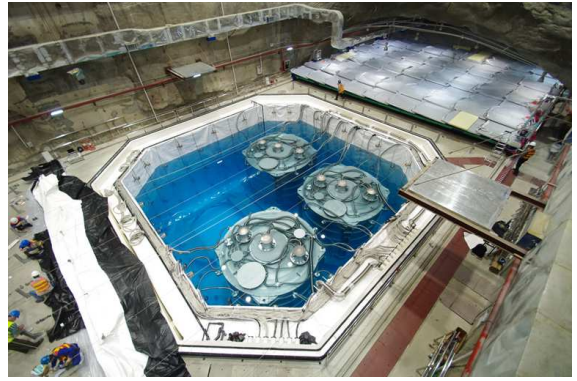}
\caption{Three ADs were deployed in the EH3.}
\label{fig:EH3}
\end{figure}

\begin{figure}
\includegraphics[width=120mm]{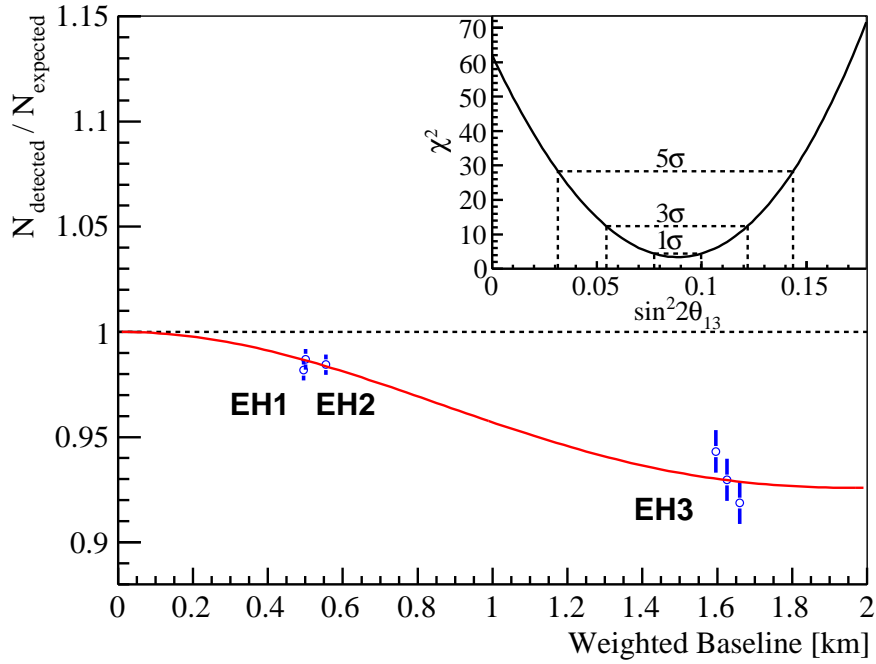}
\caption{The ratio between the measured and the expected signals in each detector vs. the 
flux-weighted average baseline, which is computed with reactor and survey data.
The expected signals have been corrected with the best fit normalization parameter 
assuming no oscillation. The error bars represent the uncorrelated uncertainties.  
The oscillation survival probability at the best-fit value is plotted as the 
smooth curve.  For clarity purpose, the AD4/6 data points are shifted by -30 and 30 m,
respectively. The $\chi^2$ value vs. the $\sin^22\theta_{13}$ value 
is shown in the inner panel.  
}
\label{fig:results}
\end{figure}

The $^9$Li/$^8$He and the fast neutrons are two major IBD contamination 
caused by cosmic muons. The nucleus of $^9$Li or $^8$He are 
produced from the carbon nucleus when cosmic muons pass through the 
liquid scintillator. The $^9$Li and $^8$He, with half lifetime of 257 ms 
and 172 ms, respectively, are both long lifetime beta emitters. 
They would undergo a beta decay providing a prompt-like signal. 
The daughter nuclei could then undergo a spontaneous fission, 
with a neutron emission in the final state resulting in a delay-like 
signal. Such a pair of correlated prompt-like and delay-like signal 
would mimic an IBD event. The contamination of $^9$Li/$^8$He can 
be directly measured by fitting the spectrum of time between the IBD candidate
and the last tagged AD muon. The measured $^9$Li/$^8$He rates from all 
three experimental halls 
are consistent with an empirical formula of $\alpha E_{\mu}^{0.74}$ 
given the average muon energy $E_{\mu}$ in each site. Furthermore, the contamination 
are suppressed by an optimized AD shower muon cut described in the 
previous section. The remaining 
contamination is about 0.2\% (0.35\%) for the far (near) site 
with a 50\% systematic uncertainty due to the fitting procedure.

The fast neutron backgrounds are caused by the energetic neutrons 
produced inside or outside the muon veto system. 
These energetic neutrons can undergo an elastic scattering with 
protons, leaving a prompt-like signal due to the proton recoil, 
followed by the neutron thermalization and then neutron capture on Gd 
producing a delay-like signal. The energy of the proton recoil 
signals ranges from sub MeV to tens of MeV. Therefore, one can 
extrapolate the measured fast neutron's prompt energy spectrum 
above 15 MeV to the energy range of interests (0.7-12 MeV)
in order to estimate the contamination in the IBD candidates. 
A flat background spectrum is assumed, which is confirmed by the
spectrum of fast neutron events with muon tagging from 
water pools and RPC. The systematic uncertainty is constrained to 
about 30\%. The contaminations due to fast neutron are estimated 
to be about 0.07\% (0.12\%) for the far (near) site.

The last contamination is the $^{13}$C($\alpha$,n)$^{16}$O background
caused by radioactivity inside ADs. The contamination is determined 
by Monte Carlo with measured alpha-decay rates. We identified four sources of 
alpha decays: the $^{210}$Po events and the decay chains from $^{238}$U, 
$^{232}$Th, and $^{227}$Ac. The backgrounds are calculated to be about 
0.05\% (0.01\%) for the far (near) site with a 50\% systematic uncertainty. 
Altogether, the total backgrounds in the IBD sample are thus determined to be 
$5\pm0.4$\% and $2\pm0.2$\% for the far site and near sites, respectively. 

\section{Oscillation Analysis}

\begin{figure}[]
\centering
\includegraphics[width=75mm]{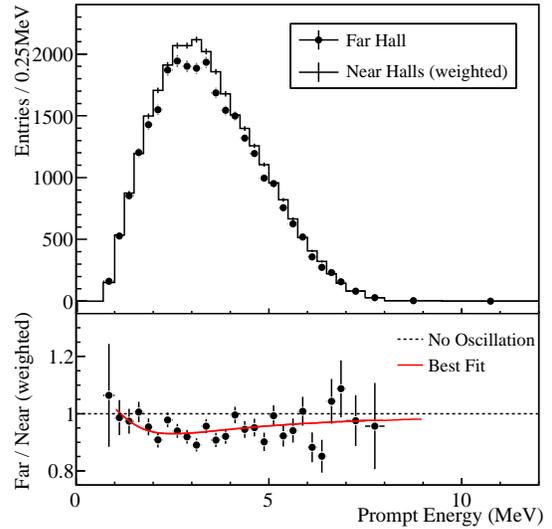}
\caption{Top panel: Measured prompt energy spectrum at the far site
(sum of three ADs) is compared with the predicted spectrum based on the 
measurements at the two near sites assuming no oscillation. Backgrounds 
are subtracted. Only statistical uncertainties are shown. 
Bottom panel: The ratio of the measured over the expected (no-oscillation) spectrum.
The solid curve is the expected ratio vs. the true prompt energy 
with the neutrino oscillation for $\sin^22\theta_{13}=0.089$.}
\label{fig:spec}
\end{figure}

The oscillation analysis includes data of
139 days (Dec. 24th 2011 - May 11th, 2012) with six ADs. Fig.~\ref{fig:EH3}
shows a picture of the three ADs in the EH3. The analysis procedure is 
described in details in Ref.~\cite{dyb_prl,dyb_cpc}. Since different 
experimental halls have different mountain overburdens, they also 
observe different muon rates. The average muon veto efficiency for six ADs 
in the three experimental halls are 0.8231, 0.8198 (EH1), 0.8576 (EH2), and 0.9813, 0.9813, 
0.9810 (EH3). Besides the background-related uncertainties, 
the largest uncorrelated detector-related uncertainty (0.12\%) 
is due to the 6 MeV energy cut in selecting delay signals. 
Other sizable uncorrelated detector-related uncertainties include 
0.1\% from the neutron Gd capture ratio, 0.03\% from the 
number of target protons, and 0.02\% from the spill-in 
effect. (The spill-in effect refers to that the IBD neutrons generated outside but captured 
inside the target GD-LS region outnumber the IBD neutrons generated inside but 
captured outside the GD-LS region. The reason for
such imbalance is that thermal neutrons 
have a larger cross section to be captured on the Gd than the proton.)
With the ratio method, the correlated detector-related 
uncertainty (about 1.9\% in total) has negligible effects 
on the oscillation analysis. The same applies to the correlated reactor-related 
uncertainty. The uncorrelated reactor-related uncertainties include 0.5\% from the received thermal power data, 
0.6\% from the calculated fission fractions,
and 0.3\% from the antineutrinos produced by the spent fuel. 
The total uncorrelated reactor-related uncertainty is 0.8\%, which is 
further suppressed by about a factor of 20 in the oscillation analysis due to the multiple 
core/reactors configuration at Daya Bay. 

The final antineutrino rates per day in the six ADs of three experimental halls, after 
corrections of the livetime, the veto efficiencies, and the background, 
are 662.47$\pm$3.00, 670.87$\pm$3.01 (EH1), 613.53$\pm$2.69 (EH2), and 77.57$\pm$0.85, 
76.62$\pm$0.85, 74.97$\pm$0.84 (EH3). In the same experimental hall, the differences 
of AD rates stemming from ADs' different locations are consistent with expectation within 
statistical uncertainties.  The deficit of antineutrino rate at the far site 
is quantified by the ratio of the measured far-hall IBD rate over 
the expected rate, which is calculated with 
the measured  IBD rates of the near detectors assuming no oscillation. 
The resulting deficit is  0.056 $\pm$ 0.007 (stat.) $\pm$ 0.003 (sys.). 

The chi-square method with pull terms is used to extract $\sin^22\theta_{13}$ within
the standard 3-flavor oscillation model, in which the disappearance probability 
of the electron antineutrino is written as:
\begin{eqnarray}
P(\bar{\nu_e}\rightarrow \bar{\nu_e}) &=& 1- \sin^2 2\theta_{13} \cos^2\theta_{12}\sin^2 \Delta_{31} \nonumber \\ 
& &	- \sin^2 2\theta_{13}\sin^2\theta_{12}\sin^2{\Delta_{32}} \nonumber \\
& & -\cos^4\theta_{13}\sin^2 2\theta_{12} \sin^2 \Delta_{21},
\end{eqnarray}
with $\Delta_{ij} \equiv |\Delta_{ij}|= 1.27 |\Delta m^2_{ij}| \frac{L(m)}{E(MeV)}$.  
In this framework, the uncertainties from the backgrounds, detectors, reactor fluxes, and 
the oscillation parameters are taken into account properly 
in a consistent manner. The $\sin^22\theta_{13}$ is determined to be
0.089$\pm$0.011. Fig.~\ref{fig:results} shows the ratios of measured over expected 
IBD rates vs. weighted baseline for all ADs. The data are compared with the 
expected oscillation curve (red-solid line). Our improved measurement 
disfavors the $\sin^22\theta_{13}=0$ hypothesis at a 7.7 standard deviations. 

\section{Conclusion and Outlook}

\begin{figure}[]
\centering
\includegraphics[width=75mm]{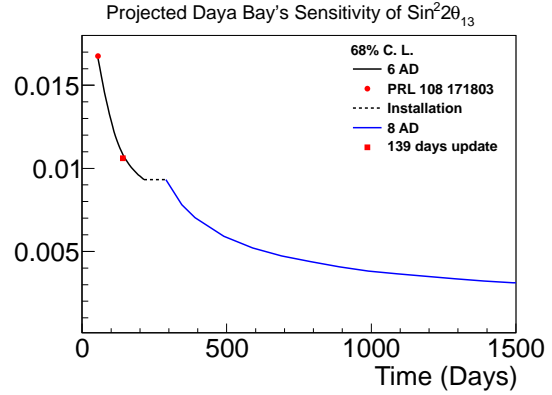}
\caption{Projected $\sin^22\theta_{13}$'s sensitivity in terms of half 
of the 68\% 
confidence interval of Daya Bay vs. the running time. 
The published and improved Daya Bay results are shown as red dots. 
The period of 6-AD data taking (black) is separated from the period 
of 8-AD data taking (blue) by an installation/calibration period (dashed). 
}
\label{fig:outlook}
\end{figure}

In this talk, we reported an improved measurement of electron 
antineutrino disappearance at Daya Bay. The new results with 2.5 
times more data confirms the previously published results~\cite{dyb_prl}, 
and improves the precision of $\sin^22\theta_{13}$. Currently, 
the total uncertainty is still dominated by the statistical 
uncertainty, which is about a factor of 2 larger than the systematic 
uncertainty. Fig.~\ref{fig:outlook} shows the projected sensitivity 
of $\sin^22\theta_{13}$ in terms of half of the 68\% confidence 
interval with respect to the running time. With the full 8-AD 
configuration, we expect to achieve a $\sim5$\% measurement of 
$\sin^22\theta_{13}$ in about 3 years. Furthermore, the Daya Bay 
experiment also has the potential to measure the effective 
squared-mass difference $\Delta m^2_{ee}$, which is a 
combination of $\Delta m^2_{31}$ and $\Delta m^2_{32}$, 
through the measurement of IBD energy spectrum distortion. Due to 
the short baseline $<2$ km, the measurement of $\Delta m^2_{ee}$ 
is not sensitive to the neutrino mass hierarchy (sign of $\Delta m^2_{32}$)~\cite{IBDMH}. 
In addition, the high statistics IBD samples from Daya Bay 
would provide the most precise measurement of the antineutrino 
energy spectrum, which is essential for the future measurements with 
reactor antineutrinos. 

With the global efforts led by the Daya Bay experiment, 
the $\sin^22\theta_{13}$ is found to be around 0.09. Such a large value 
of $\sin^22\theta_{13}$ opens doors to two of the remaining unknown 
parameters in the neutrino sector, the CP phase $\delta$ in the leptonic 
sector and the neutrino Mass Hierarchy~\footnote{A Bayesian approach in 
presenting results/sensitivities of the neutrino mass hierarchy has been 
proposed in Ref.~\cite{StatMH}.}. In particular, the long baseline 
experiments~\cite{t2k,nova,LBNE,hyperk} can provide essential information for both 
parameters through the (anti)$\nu_{e}$ appearance from a (anti)$\nu_{\mu}$ beam. 
Meanwhile, the possibility of utilizing a medium baseline ($\sim$60 km) reactor neutrino 
experiment to determine the neutrino mass hierarchy is also intensively 
discussed~\cite{Petcov,Learned,zhan,zhan1,IBDMH,ihep}. We therefore expect a new era
of discovery in the next couple of decades.


\begin{theacknowledgments}
 We would like to thank the support from Caltech and the National 
Science Foundation. The Daya Bay experiment is supported in part by the Ministry
of Science and Technology of China, the United States
Department of Energy, the Chinese Academy of Sciences, the
National Natural Science Foundation of China, the Guangdong
provincial government, the Shenzhen municipal government,
the China Guangdong Nuclear Power Group, Shanghai
Laboratory for Particle Physics and Cosmology, the Research
Grants Council of the Hong Kong Special Administrative Region
of China, University Development Fund of The University
of Hong Kong, the MOE program for Research of Excellence
at National Taiwan University, National Chiao-Tung
University, and NSC fund support from Taiwan, the U.S. National
Science Foundation, the Alfred P. Sloan Foundation, the
Ministry of Education, Youth and Sports of the Czech Republic,
the Czech Science Foundation, and the Joint Institute of
Nuclear Research in Dubna, Russia. We thank Yellow River
Engineering Consulting Co., Ltd. and China railway 15th Bureau
Group Co., Ltd. for building the underground laboratory.
We are grateful for the ongoing cooperation from the China
Guangdong Nuclear Power Group and China Light \& Power Company. 
\end{theacknowledgments}



\bibliographystyle{aipproc}   

\begin{thebibliography}{9}
\bibitem{helicity} M. Goldhaber, L. Grodzins, and A. W. Sunyar, Phys. Rev. {\bf 109}, 1015 (1958).
\bibitem{mckeown_review} 
R. D. McKeown and P. Vogel, Phys. Rep. {\bf 394}, 315 (2004).
\bibitem{ponte1} B. Pontecorvo, Sov. Phys. JETP {\bf 6}, 429 (1957).
\bibitem{ponte2} B. Pontecorvo, Sov. Phys. JETP {\bf 26}, 984 (1968).
\bibitem{Maki} Z. Maki, M. Nakagawa, and S. Sakata, Prog. Theor. Phys. {\bf 28}, 870 (1962).
\bibitem{chooz1} M. Apollonio {\it et al.} [CHOOZ Collaboration], Phys. Lett. {\bf B466}, 415 (1999).
\bibitem{chooz2} M. Apollonio {\it et al.} [CHOOZ Collaboration], Eur. Phys. J. {\bf C27}, 331 (2003).
\bibitem{Fogli} G. Fogli {\it et al.}, Phys. Rev. Lett. {\bf 101}, 141801 (2008).
\bibitem{T2k} K. Abe {\it et al.} [T2K Collaboration], Phys. Rev. Lett. {\bf 107}, 041801 (2011).
\bibitem{MINOS} P. Adamson {\it et al.} [MINOS Collaboration], Phys. Rev. Lett. {\bf 107}, 181802 (2011).
\bibitem{Dchooz} Y. Abe {\it et al.} [Double Chooz Collaboration], Phys. Rev. Lett. {\bf 108}, 131801 (2012).
\bibitem{ratio} L. A. Mikaelyan and V. V. Sinev, Phys. Atom. Nucl. {\bf 63}, 1002 (2000).
\bibitem{dyb_prl}
F. P. An {\it et al.} [Daya Bay Collaboration], Phys. Rev. Lett. {\bf 108}, 171803 (2012).
\bibitem{reno} J. K. Ahn {\it et al.} [RENO Collaboration], Phys. Rev. Lett. {\bf 108}, 191802 (2012).
\bibitem{dyb_cpc} F. P. An {\it et al.}[Daya Bay Collaboration], arXiv:1210.6327 (2012), submitted to Chinese Physics C.
\bibitem{dyb_proposal} Daya Bay Collaboration, arXiv:hep-ex/0701029 (2006). 
\bibitem{dyb_nim}
F. P. An {\it et al.} [Daya Bay Collaboration],  Nucl. Inst. Method {\bf A685}, 78 (2012).
\bibitem{IBDMH} X. Qian {\it et al.}, arXiv:1208.1551 (2012), submitted to Phys. Rev. D.
\bibitem{StatMH} X. Qian {\it et al.}, arXiv:1210.3651 (2012), submitted to Phys. Rev. D.
\bibitem{t2k} K. Abe {\it et al.} [T2K Collaboration] Nucl. Inst. Method {\bf A659}, 106 (2011).
\bibitem{nova} D. Ayres {\it et al.} [NO$\nu$A Collaboration] The NO$\nu$A Technical Design Report. 
	FERMILAB-DESIGN-2007-01 (2007). 
\bibitem{LBNE} T. Akiri {\it et al.} [LBNE Collaboration], arXiv:1110.6249 (2011).
\bibitem{hyperk} K. Abe {\it et al.} [Hyper-K Collaboration], arXiv:1109.3262 (2011).
\bibitem{Petcov} S. Choubey, S. T. Petcov, and M. Piai, Phys. Rev. {\bf D68}, 113006 (2003).
\bibitem{Learned} J. G. Learned {\it et al.} Phys. Rev. {\bf D78}, 071302R (2008).
\bibitem{zhan} L. Zhan {\it et al.} Phys. Rev. {\bf D78}, 111103R (2008).
\bibitem{zhan1} L. Zhan {\it et al.} Phys. Rev. {\bf D79}, 073007 (2009).
\bibitem{ihep} E. Ciuffoli, J. Evaslin, and X. M. Zhang, arXiv:1209.2227 (2012). 
\end{thebibliography}



\end{document}